\documentstyle[sprocl]{article}

\def\Journal#1#2#3#4{{#1} {\bf #2}, #3 (#4)}

\def\NPB{{\em Nucl. Phys.} B}
\def\PLB{{\em Phys. Lett.}  B}
\def\PRL{\em Phys. Rev. Lett.}
\def\PRD{{\em Phys. Rev.} D}

\def\be{\begin{equation}}
\def\ee{\end{equation}}
\def\bea{\begin{eqnarray}}
\def\eea{\end{eqnarray}}

\begin{document}

\title{EXCITED HEAVY BARYON MASSES FROM THE $1/N_c$ EXPANSION OF HQET}

\author{JONG-PHIL LEE}

\address{Department of Physics, Seoul National University, Seoul 
151-742, Korea\\E-mail: jplee@phya.snu.ac.kr}

\author{CHUN LIU\footnote{Speaker}}

\address{Institute of Theoretical Physics, CAS, Beijing 100080, China
\\E-mail: liuc@itp.ac.cn}

\author{H. S. SONG}

\address{Department of Physics, Seoul National University, Seoul 
151-742, Korea\\E-mail: hssong@physs.snu.ac.kr}  

\maketitle\abstracts{ The mass spectra of the $L=1$ orbitally excited 
heavy baryons with light quarks in both the spin-flavor symmetric and 
the mixed representations are studied by the $1/N_c$ expansion method 
in the framework of the heavy quark effective theory.  The mixing effect 
between the baryons in the two representations is also considered.  The 
general pattern of the spectrum is predicted which will be verified by
the experiments in the near future.}

\section{Introduction}
Experimentally, a lot of data for orbitally excited heavy baryons 
have been accumulating \cite{1}.  The following charmed baryon states 
have been found, $\Lambda_c(2593)^+$ with $I(J^P)=0(\frac{1}{2}^-)$ 
which is denoted as $\Lambda_{c1}(\frac{1}{2}^-)^+$, 
$\Lambda_c(2625)^+$ with $I(J^P)=0(?)$ denoted as 
$\Lambda_{c1}(\frac{3}{2}^-)^+$, where $?$ is $\frac{3}{2}^-$ from the 
quark model, and its strange analogues $\Xi_{c1}(\frac{3}{2}^-)$.  

Theoretical understanding of these baryons is necessary.  The heavy 
quark effective theory (HQET) \cite{3} provides a systematic way to 
investigate hadrons containing a single heavy quark.  To obtain 
detailed prediction, however, some non-perturbative QCD methods have 
to be used, such as lattice simulation, $1/N_c$ expansion 
\cite{4,5}, chiral Lagrangian and QCD sum rules.  In this talk we 
report the application of the $1/N_c$ expansion method \cite{6,7}.  

Let us first make a brief review of the HQET \cite{3}.  It is an 
effective field theory of QCD for heavy hadrons.  In the limit 
$m_Q/\Lambda_{\rm QCD}\to\infty$, the heavy quark spin-flavor 
symmetry (HQS) is explicit.  The 4-velocity $v$ of the heavy quark 
becomes a good quantum number.  Because $m_Q$ is unusable, it can 
be removed by redefining the heavy quark field:
\be
h_v = e^{im_Qv\cdot x}Q\;.
\label{eq:1}
\ee
The effective Lagrangian is then 
\be
{\cal L}_{\rm eff}=\bar{h}_vv\cdot Dh_v+O(1/m_Q)\;.
\label{eq:2}
\ee
The hadron mass is expanded as 
\be
\bar{\Lambda}_H+O(1/m_Q) 
\label{eq:3}
\ee
in the effective theory, that is 
\be
M_H = m_Q+\bar{\Lambda}_H+O(1/m_Q)\;.
\label{eq:4}
\ee

Second, let us come to the $1/N_c$ expansion \cite{4}.  This is a 
non-perturbative method of QCD.  The idea is to extract out the 
non-perturbative information of $SU(N_c)$ gauge theory by taking 
$N_c\to\infty$.  The $N_c$ counting rules are given as follows.  
The interaction vertex is $g_s/\sqrt{N_c}$; the quark propagator 
keeps unchanged; and the gluon propagator is represented by double 
lines, one for quarks and the other for anti-quarks.  

For the mesons, the non-perturbative properties can be observed 
from the analysis of the planar diagrams.  The large $N_c$ limit 
is quite successful.  Because the meson decay amplitude 
$\sim 1/\sqrt{N_c}$, mesons and glue states are free and stable.  
This agrees qualitatively with color confinement.  Another 
example is the explanation of the Zweig's rule.  

For the baryons, the diagrammatic method does not work.  The 
Hartree approximation can be adopted.  The observation is that in 
the $N_c\to\infty$ limit, interaction between any pair of quarks is 
negligible ($\sim 1/N_c$); the total potential on an individual quark, 
which is $\sim 1$, is a sum of many small terms, therefore it can be 
regarded as the background potential or a c-number potential.  For 
ground state baryons, the many-body wave function is written as 
\be
\Psi(x_1, ..., x_{N_c}, t) = \sum^{N_c}_1\phi (x, t)\;,
\label{eq:5}
\ee
where $\phi (x, t)$ is one-body state.  Interesting results for large 
$N_c$ baryons can be obtained.  The baryon-(anti-)baryon interaction 
is order $N_c$.  And the baryon-meson interaction is order 1.  However, 
the Hartree potential is not known, in which the two-body, three-body, 
and many-body interactions are the same important.  One conjecture is 
that baryons are solitons of the mesonic theory, skyremions \cite{8}!  

Something more can be said about the large $N_c$ baryons \cite{5}.  For 
the ground state baryons, there is a contracted SU(2$N_f$) light quark 
spin-flavor symmetry (LQS) in the large $N_c$ limit.  This was first 
obtained from the chiral perturbation theory of baryon-pion  
interactions in deriving the consistent conditions for the coupling 
constants in the large $N_c$ limit.  It can be also understood in the 
Hartree picture.  This makes a $1/N_c$ expansion based on the 
spin-flavor structure practically possible for the baryons.  Many 
quantitative predictions and further extensions of the above result 
have been made \cite{9}.  In fact, this expansion is another scheme 
of the $1/N_c$ expansion.  This can be simply seen from considering 
the masses of the non-strange baryons, 
\begin{equation}
\begin{array}{rcl}
M_H &=& \displaystyle N_c\Lambda_{\rm QCD}+O(1)+O(\frac{1}{N_c})+...
\\[4pt]
&=& \displaystyle N_c\tilde{\Lambda}_{\rm QCD}+ c_1\frac{S^2}{N_c} 
        +c_2\frac{(S^2)^2}{N_c^2}+...
\;,
\end{array}\label{eq:6}
\end{equation}
where the first line is the ordinary $1/N_c$ expansion, and the 
second one the expansion based on the spin-flavor structure.  Of 
course, in the $N_c\to\infty$ limit, 
$M_H=\bar{\Lambda}_H=N_c\Lambda_{\rm QCD}=N_c\tilde{\Lambda}_{\rm QCD}
=m_{proton}$ 
which is not so useful.  
 
\section{Excited Heavy Baryons in the $1/N_c$ Expansion}
For the charmed baryons like $\Lambda_{c1}(\frac{1}{2}^-)^+$, 
$\Lambda_{c1}(\frac{3}{2}^-)^+$, $\Xi_{c1}(\frac{3}{2}^-)^+$, we 
analyze their masses $\bar{\Lambda}_H$ in the $1/N_c$ expansion.  The 
classification of them is according to the angular momentum $J$, the 
isospin $I$ and the total angular momentum of the light degrees of 
freedom $J^l$ which becomes a good quantum number due to HQS.  In this 
case, the excited hadron spectrum shows the degeneracy of pair of 
states which are related to each other by HQS.  Constituently there are 
two ways for the $L=1$ excitation.  One is that the heavy quark is 
excited; the other is that one light quark is excited.  
Correspondingly under the LQS, the $N_c-1$ light quarks are in the 
symmetric and mixed representation, respectively.  

\subsection{Symmetric Representation}\label{subsec:symmetric}
In the symmetric representation, the picture for the light quarks is
essentially the same as that of the ground state heavy baryons.  The
spin-flavor decomposition rule is $I=S^l$ for the non-strange baryons, 
where $S^l$ is the total spin of the light quark system.  Note that 
the light quark system as a whole has $L=1$ orbital angular momentum.  
All possible states of excited heavy baryons are listed in 
Table~\ref{tab:sym}.  

\begin{table}[t]
\caption{Excited heavy baryon states of the symmetric representation 
of $N_c-1$ light quarks. \label{tab:sym}}
\vspace{0.2cm}
\begin{center}
\footnotesize
\begin{tabular}{|c|c|c|}
\hline
{$(J,I)$} &$(J^l,S^l)$&$\bar{\Lambda}_H^0$\\
\hline
$(1/2,0)$ & $(1,0)$     & $N_cc_0+2c_1$\\
$(3/2,0)$ & $(1,0)$     & $N_cc_0+2c_1$\\
$(1/2,1)$ & $(0,1)$     & $N_cc_0-2c_1+\frac{2c_2}{N_c}$\\
$(1/2,1)$ & $(1,1)$     & $N_cc_0+\frac{2c_2}{N_c}$\\
$(3/2,1)$ & $(1,1)$     & $N_cc_0+\frac{2c_2}{N_c}$\\
$(3/2,1)$ & $(2,1)$     & $N_cc_0+4c_1+\frac{2c_2}{N_c}$\\
$(5/2,1)$ & $(2,1)$     & $N_cc_0+4c_1+\frac{2c_2}{N_c}$\\
\hline
\end{tabular}
\end{center}
\end{table}
 
In the Hartree--Fock picture of the baryons, the $N_c$ counting rules 
require us to include many-body interactions in the analysis.  
However, a large part of these interactions are spin-flavor irrelevant.  
Namely this part contributes in the order $N_c\Lambda_{\rm QCD}$ 
universally to all the baryons with different spin-flavor structure in 
Table~\ref{tab:sym}.  The mass splittings among the baryons can be 
obtained.  For the purely light quark contribution to 
$\bar{\Lambda}_H$, the $1/N_c$ analysis goes the same as that to the 
ground state heavy baryons.  There is LQS at the leading order of the 
$1/N_c$ expansion.  The mass splitting due to the violation of LQS 
started from ${S^l}^2/N_c$.  However, different from the ground state 
baryons, formally the orbital angular momentum of the heavy quark has 
more dominant contribution to $\bar{\Lambda}_H$ than $O(1/N_c)$.  This 
is because of the orbital-light-quark-spin interactions.  After 
summing up all the relevant many-body interactions, this order $O(1)$ 
contribution is 
$\displaystyle\vec{L}\cdot\vec{S^l}f(\frac{{S^l}^2}{N_c^2})$, where 
$f$ is a general function which can be Taylor expanded.  The mass 
$\bar{\Lambda}_H$ can be written simply as
\be
\bar{\Lambda}^0_H=N_c\tilde{c}_0+\tilde{c}_1\vec{L}\cdot\vec{S^l}
     +O\left(\frac{1}{N_c}\right)\,,
\label{7}
\ee
where coefficients $\tilde{c}_i\sim\Lambda_{\rm QCD}$ ($i=0,1$). There 
should be also a term proportional to $L^2$ in the above equation, 
which gives constant contribution to $\bar{\Lambda}^0_H$ for a given 
light quark representation, and therefore has been absorbed into the 
leading term.  The term $\vec{L}\cdot\vec{S^l}$ can be rewritten as 
${J^l}^2-{S^l}^2$ with $\vec{J^l}$ being defined as 
$\vec{J^l}=\vec{L}+\vec{S^l}$.  Therefore 
\be 
\bar{\Lambda}^0_H=N_cc_0+c_1({J^l}^2-{S^l}^2)
            +O\left(\frac{1}{N_c}\right)\,,
\label{8}
\ee
where coefficients $c_i\sim\Lambda_{\rm QCD}$.

\subsection{Mixed Representation}\label{subsec:mixed}
In the mixed representation, all states are listed in 
Table~\ref{tab:mx}.  Again, $\bar{\Lambda}_{H'}$ is trivially 
$N_c\Lambda_{\rm QCD}$ at the leading order of the $1/N_c$ expansion.  
The spin-flavor dependence, however is more complicated.  For the 
spectrum of excited light baryons, see ref. \cite{9}.

\begin{table}[t]
\caption{Excited heavy baryon states of the mixed representation of 
$N_c-1$ light quarks. \label{tab:mx}}
\vspace{0.2cm}
\begin{center}
\footnotesize
\begin{tabular}{|c|c|c|}
\hline
{$(J,I)$} &$(J^l,S^l)$&$\bar{\Lambda}_{H'}^0$\\
\hline
$(1/2,0)$ & $(0,1)$ &
$-2c_{LS}+\frac{1}{18}c_T-\frac{1}{3}\bar{c}_1+\frac{1}{2}\bar{c}_2$\\ 
$(1/2,0)$ & $(1,1)$ &
$-c_{LS}+\frac{1}{18}c_T-\frac{1}{6}\bar{c}_1-\frac{1}{4}\bar{c}_2$\\
$(3/2,0)$ & $(1,1)$ &
$-c_{LS}+\frac{1}{18}c_T-\frac{1}{6}\bar{c}_1-\frac{1}{4}\bar{c}_2$\\
$(3/2,0)$ & $(2,1)$ &
$9c_{LS}+\frac{1}{18}c_T+\frac{1}{6}\bar{c}_1+\frac{1}{20}\bar{c}_2$\\
$(5/2,0)$ & $(2,1)$ &
$9c_{LS}+\frac{1}{18}c_T+\frac{1}{6}\bar{c}_1+\frac{1}{20}\bar{c}_2$\\
$(1/2,1)$ & $(1,0)$ & $\frac{1}{18}c_T$\\
$(3/2,1)$ & $(1,0)$ & $\frac{1}{18}c_T$\\
\hline
\end{tabular}
\end{center}
\end{table}

The many-body Hamiltonians related to the spin-flavor structure which 
involve orbital angular momentum $L$ give $O(1)$ contribution.  We use 
the following operators which were used in \cite{9} to analyze 
$\bar{\Lambda}_H$,
\bea
H_{LS}&\propto&\hat{a}^{\dagger}~\vec{L}\cdot\vec{\sigma}~\hat{a}\nonumber\\
H_T&\propto&\frac{1}{N_c}G^{ia}G_{ia}\nonumber\\
H_1&\propto&\frac{1}{N_c}\hat{a}^{\dagger}~L^i\otimes\tau^a~\hat{a}~G_{ia}
\nonumber\\
H_2&\propto&\frac{1}{N_c}\hat{a}^{\dagger}\{L^i,L^j\}\otimes\sigma_i
\otimes\tau^a~\hat{a}~G_{ja}~.
\label{9}
\eea
The first one $H_{LS}$ is one-body Hamiltonian, while the others are 
two-body Hamiltonians.  $G_{ia}$ are the generators of the spin-flavor 
symmetry group SU(4), given by
\be
G^{ia}=\hat{a}^{\dagger}~\sigma^i\otimes\tau^a~\hat{a}
\label{10}
\ee
with $\sigma^i$ and $\tau^a$ being the spin and isospin matrices, 
respectively.  Such structure gives coherent addition over $N_c-2$ core 
quarks.  The first $G^{ia}$ in $H_T$ acts on the excited quark, the other 
$G_{ia}$'s on the $N_c-2$ unexcited light quarks, namely the core quarks.
In our case, all the operators must be understood as the ones acting on 
the light degrees of freedom.  Note that the higher order many-body 
Hamiltonian which contains more factor of $G_{ia}$ can be reduced to 
those given in Eq.(\ref{9}).

The contributions to the baryon masses due to these Hamiltonians are 
obtained by calculating the baryonic matrix elements.  The matrix 
elements of these operators between the states of light quarks which 
specify the states of excited heavy baryons are given as follows,
\begin{eqnarray}
&&\langle I_c=\frac{1}{2};~I ~I_3;~S^{l\prime}~S^{l\prime}_3, 
  ~l=1~m^{\prime}~|~H_T~|~
  I_c=\frac{1}{2};~I~I_3;~S^l~S^l_3,~l=1~m\rangle\nonumber\\
&=&2c_T\delta_{S^{l\prime},S^l}\delta_{S^{l\prime}_3,S^l_3}
  \delta_{m,m^\prime}(-1)^{1-S^l-I}
  \left\{\begin{array}{ccc}
  S^l&\frac{1}{2}&\frac{1}{2}\\
  1&\frac{1}{2}&\frac{1}{2}\end{array}\right\}
  \left\{\begin{array}{ccc}
  \frac{1}{2}&1&\frac{1}{2}\\
  \frac{1}{2}&I&\frac{1}{2}\end{array}\right\}~,\nonumber\\
\\
&&\langle I_c=\frac{1}{2};~I~I_3;~l=1,~S^{l\prime},~J^l~J^l_3~|~H_{LS}~|
  ~I_c=\frac{1}{2};~I~I_3;~l=1,~S^l,~J^l~J^l_3\rangle\nonumber\\
&=&c_{LS}(-1)^{S^l-S^{l\prime}}\sqrt{(2S^l+1)(2S^{l\prime}+1)}
  \sum_{j=\frac{1}{2},\frac{3}{2}}(2j+1)\{j(j+1)-2-3/4\}\nonumber\\
&&\left\{\begin{array}{ccc}
  \frac{1}{2}&\frac{1}{2}&S^l\\
  1&J^l&j\end{array}\right\}
  \left\{\begin{array}{ccc}
  \frac{1}{2}&\frac{1}{2}&S^{l\prime}\\
  1&J^l&j\end{array}\right\}~,\nonumber\\
\\
&&\langle I_c=\frac{1}{2};~I~I_3;~l=1,~S^{l\prime},~J^l~J^l_3~|~H_1~|
  ~I_c=\frac{1}{2};~I~I_3;~l=1,~S^l,~J^l~J^l_3\rangle\nonumber\\
&=&6\bar{c}_1(-1)^{I-J^l+S^l-S^{l\prime}-1}
  \sqrt{(2S^l+1)(2S^{l\prime}+1)}
  \left\{\begin{array}{ccc}
  \frac{1}{2}&1&\frac{1}{2}\\
  \frac{1}{2}&I&\frac{1}{2}\end{array}\right\}
  \left\{\begin{array}{ccc}
  S^l&1&S^{l\prime}\\
  \frac{1}{2}&\frac{1}{2}&\frac{1}{2}\end{array}\right\}
  \left\{\begin{array}{ccc}
  1&1&1\\
  S^{l\prime}&J^l&S^l\end{array}\right\}~,\nonumber\\
\\
&&\langle I_c=\frac{1}{2};~I~I_3;~l=1,~S^{l\prime},~J^l~J^l_3~|~H_2~|
  ~I_c=\frac{1}{2};~I~I_3;~l=1,~S^l,~J^l~J^l_3\rangle\nonumber\\
&=&3\bar{c}_2(-1)^{1+J^l+I+S^{l\prime}+2S^l}
  \sqrt{(2S^l+1)(2S^{l\prime}+1)}
  \left\{\begin{array}{ccc}
  \frac{1}{2}&1&\frac{1}{2}\\
  \frac{1}{2}&I&\frac{1}{2}\end{array}\right\}
  \left\{\begin{array}{ccc}
  2&1&1\\
  J^l&S^{l\prime}&S^l\end{array}\right\}
  \left\{\begin{array}{ccc}
  S^{l\prime}&S^l&2\\
  \frac{1}{2}&\frac{1}{2}&1\\
  \frac{1}{2}&\frac{1}{2}&1\end{array}\right\}~,
\label{11-14}
\end{eqnarray}
where $I_c$ is the isospin of the core quarks.  In real world ($N_c=3$), 
there is only one quark in the core so $I_c$ always equals 
$\frac{1}{2}$.  With these matrix elements, we can express the excited 
heavy baryon mass up to the zeroth order of $1/N_c$:
\be
\bar{\Lambda}_{H'}=N_c\bar{c}_0 +\langle H_{LS}\rangle+\langle H_T\rangle+
  \sum_{i=1}^{2}\langle H_i\rangle~,
\label{15}
\ee
where $c_{LS}$, $c_T$, and $\bar{c}_i$'s are the coefficients 
$\sim \Lambda_{\rm QCD}$.

\subsection{Mixing}\label{subsec:mixing}
It is necessary to consider the mixing between the baryons with light 
quarks in the spin-flavor symmetric and mixed representations.  When they 
have same good quantum numbers of ($J$, $I$, $J^l$), there is no physical 
way to distinguish them.  This consideration will give the physical spectrum.  
Because of the light quark spin-flavor symmetry at the leading order of 
$1/N_c$ expansion, the baryons with same ($J$, $I$, $J^l$) quantum numbers 
but in different representations do not mix. ($S^l$ is a good quantum number 
in $N_c\to \infty$.) \cite{2}.  The mixing occurs at the sub-leading order.  
The classification of baryons by the spin-flavor symmetry  is therefore 
physical at the leading order.  For the physical spectrum, the mixing 
results in a deviation from $\bar{\Lambda}^0_H$.  By denoting the mixing 
mass as $\tilde{m}$ which is of $O(1)$, the mass matrix for the baryons with 
same ($J, I, J^l$) is written as
\be
\left(\begin{array}{cc}
\bar{\Lambda}^0_H&\tilde{m}           \\
\tilde{m}        &\bar{\Lambda}^0_{H'}\\
\end{array}
\right)
\,.
\label{16}
\ee
The mass difference $\bar{\Lambda}^0_H-\bar{\Lambda}^0_{H'}$ is $O(1)$.  
Taking $\tilde{m}<\bar{\Lambda}^0_{H'}-\bar{\Lambda}^0_H$ for illustration, 
the physical masses are corrected to be
\begin{equation}
\begin{array}{lll}
\bar{\Lambda}_H   &\simeq&\displaystyle\bar{\Lambda}^0_H-\frac{\tilde{m}^2}
{\bar{\Lambda}^0_{H'}-\bar{\Lambda}^0_H}~,\\[3mm]
\bar{\Lambda}_{H'}&\simeq&\displaystyle\bar{\Lambda}^0_{H'}
+\frac{\tilde{m}^2}{\bar{\Lambda}^0_{H'}-\bar{\Lambda}^0_H}~.
\label{17}
\end{array}
\end{equation}
The $1/N_c$ expansion of $\tilde{m}$ is parameterized as
\be
\tilde{m}=\tilde{m}_0+O(1/N_c)\,,
\label{18}
\ee
where $\tilde{m}_0$ is universal due to LQS.  To the order of $O(1)$, the 
spectrum is,
\begin{equation}
\begin{array}{lll}
\displaystyle\bar{\Lambda}_{(\frac{1}{2}(\frac{3}{2}), 0, 1)}&=&\displaystyle
N_cc_0+2c_1-\frac{\tilde{m}_0^2}
{k-c_{LS}-\frac{1}{6}\bar{c}_1-\frac{1}{4}\bar{c}_2-2c_1}~,\\[3mm]
\displaystyle\bar{\Lambda}_{(\frac{1}{2}, 1, 0)}      &=&N_cc_0-2c_1~,\\[3mm]
\displaystyle\bar{\Lambda}_{(\frac{1}{2}(\frac{3}{2}), 1, 1)}&=&\displaystyle
N_cc_0-\frac{\tilde{m}_0^2}{k}~,\\[3mm]
\displaystyle\bar{\Lambda}_{(\frac{3}{2}(\frac{5}{2}), 1, 2)}&=&N_cc_0+4c_1~,
\label{19}
\end{array}
\end{equation}
where $k$ is an $O(1)$ constant which is what remains after the
$\bar{\Lambda}^0_{H'}$ and $\bar{\Lambda}^0_H$ cancellation.  Note that the 
masses of the states $\displaystyle (\frac{1}{2}, 1, 0)$ and
$\displaystyle (\frac{3}{2}(\frac{5}{2}), 1, 2)$ are not affected by the
mixing, because there are no physical states with the same good quantum
numbers in the mixed representation.  From the above spectrum, we see that
$c_1>0$.  The states $\displaystyle (\frac{3}{2}(\frac{5}{2}), 1, 2)$ is
always the highest states.  They are heavier than the other states at least
by $4c_1$ through requiring the states
$\displaystyle (\frac{1}{2}(\frac{3}{2}), 0, 1)$ to be the lowest.  If
$\displaystyle 2c_1>\frac{\tilde{m}_0^2}{k}$, the requirement implies
\be
\frac{\tilde{m}_0^2}
{k-c_{LS}-\frac{1}{6}\bar{c}_1-\frac{1}{4}\bar{c}_2-2c_1}>4c_1~.
\label{20}
\ee
In this case, the spectrum pattern is
\be 
M(\frac{1}{2}(\frac{3}{2}), 0, 1)<M(\frac{1}{2}, 1, 0)< 
M(\frac{1}{2}(\frac{3}{2}), 1, 1)<M(\frac{3}{2}(\frac{5}{2}), 1, 2)\,. 
\label{21}
\ee
On the other hand, if $\displaystyle 2c_1<\frac{\tilde{m}_0^2}{k}$, the
requirement is
\be
\tilde{m}_0^2\left(\frac{1}
{k-c_{LS}-\frac{1}{6}\bar{c}_1-\frac{1}{4}\bar{c}_2-2c_1}-\frac{1}{k}\right)
>2c_1~,
\label{22}
\ee
which gives the spectrum
\be 
M(\frac{1}{2}(\frac{3}{2}), 0, 1)<M(\frac{1}{2}(\frac{3}{2}), 1, 1)<
M(\frac{1}{2}, 1, 0)<M(\frac{3}{2}(\frac{5}{2}), 1, 2)\,.
\label{23}
\ee

The experimentally found baryons $\Lambda_{c1}(\frac{1}{2})$ and 
$\Lambda_{c1}(\frac{3}{2})$ correspond to the 
$\displaystyle (\frac{1}{2}(\frac{3}{2}), 0, 1)$ states.  More data are
needed to fix the unknown parameters $c_i$'s, $\bar{c_i}$'s, $k$ and
$c_{LS}$.  In the near future, experiments will check the above predicted
spectrum.  Hopefully one of the above mass patterns will be picked out. It
will be a check for the validity of our method, if the parameters are in 
the reasonable range ($\Lambda_{\rm QCD}$) and meanwhile satisfy the 
relations given above.

\section{Summary}
In summary, we have reported the $1/N_c$ expansion method in studying the  
spectra of the $L=1$ orbitally excited heavy baryons within the framework 
of HQET.  The analysis is very simple for the baryons with light quarks 
being in the spin-flavor symmetric representation, compared to that for the 
heavy baryons with light quarks in the mixed representation.  The 
simplicity is an unique feature.  It can be seen from the point that the 
light quark system is in the ground state and it is the heavy quark that is 
orbitally excited.  However the mixing effect due to the baryon states in 
the mixed representation corrects the spectrum pattern in the sub-leading 
order of $1/N_c$ expansion.  The effect is important to get the realistic 
spectra at this order.  The general pattern of the baryon spectrum has been 
given, which will be verified by the experiments in the near future.  The 
$1/m_Q$ and SU(3) corrections have been considered in ref.~\cite{6}. Certain 
mass relations for the baryons $\Lambda_{c1}^{(*)}$, $\Sigma_{c1}^{(*)}$, 
$\Xi_{c1}^{(')(*)}$, and $\Omega_{c1}^{(*)}$ have been derived.  The same 
analysis can be applied to the bottom baryons. 


\section*{Acknowledgments}
This work was supported in part by the National Natural Science 
Foundation of China and BK21 Program of the Ministry of Education of 
Korea.


\end{document}